
\input harvmac
%
%
%
%
%
%
%
%
%
%
%
\newif\ifdraft

\noblackbox
\catcode`\@=11
\newif\iffrontpage
%
\ifx\answ\bigans
\def\titleft{\titsm}
\magnification=1200\baselineskip=15pt plus 2pt minus 1pt
%
\advance\hoffset by-0.075truein
\hsize=6.15truein\vsize=600.truept\hsbody=\hsize\hstitle=\hsize
\else\let\lr=L
\def\titleft{\titla}
\magnification=1000\baselineskip=14pt plus 2pt minus 1pt
%
\vsize=6.5truein
\hstitle=8truein\hsbody=4.75truein
\fullhsize=10truein\hsize=\hsbody
\fi
\parskip=4pt plus 10pt minus 4pt

\font\titla=cmr10 scaled\magstep3
\font\tenmss=cmss10
\font\absmss=cmss10 scaled\magstep1
\newfam\mssfam
\font\footrm=cmr8  \font\footrms=cmr5
\font\footrmss=cmr5   \font\footi=cmmi8
\font\footis=cmmi5   \font\footiss=cmmi5
\font\footsy=cmsy8   \font\footsys=cmsy5
\font\footsyss=cmsy5   \font\footbf=cmbx8
\font\footmss=cmss8
\def\footfont{\def\rm{\fam0\footrm}
\textfont0=\footrm \scriptfont0=\footrms
\scriptscriptfont0=\footrmss
\textfont1=\footi \scriptfont1=\footis
\scriptscriptfont1=\footiss
\textfont2=\footsy \scriptfont2=\footsys
\scriptscriptfont2=\footsyss
\textfont\itfam=\footi \def\it{\fam\itfam\footi}
\textfont\mssfam=\footmss \def\mss{\fam\mssfam\footmss}
\textfont\bffam=\footbf \def\bf{\fam\bffam\footbf} \rm}
\def\tenpoint{\def\rm{\fam0\tenrm}
\textfont0=\tenrm \scriptfont0=\sevenrm
\scriptscriptfont0=\fiverm
\textfont1=\teni  \scriptfont1=\seveni
\scriptscriptfont1=\fivei
\textfont2=\tensy \scriptfont2=\sevensy
\scriptscriptfont2=\fivesy
\textfont\itfam=\tenit \def\it{\fam\itfam\tenit}
\textfont\mssfam=\tenmss \def\mss{\fam\mssfam\tenmss}
\textfont\bffam=\tenbf \def\bf{\fam\bffam\tenbf} \rm}
\ifx\answ\bigans\def\abstractfont{\tenpoint}\else
\def\abstractfont{\def\rm{\fam0\absrm}
\textfont0=\absrm \scriptfont0=\absrms
\scriptscriptfont0=\absrmss
\textfont1=\absi \scriptfont1=\absis
\scriptscriptfont1=\absiss
\textfont2=\abssy \scriptfont2=\abssys
\scriptscriptfont2=\abssyss
\textfont\itfam=\bigit \def\it{\fam\itfam\bigit}
\textfont\mssfam=\absmss \def\mss{\fam\mssfam\absmss}
\textfont\bffam=\absbf \def\bf{\fam\bffam\absbf}\rm}\fi
%
\def\f@@t{\baselineskip10pt\lineskip0pt\lineskiplimit0pt
\bgroup\aftergroup\@foot\let\next}
\setbox\strutbox=\hbox{\vrule height 8.pt depth 3.5pt width\z@}
\def\vfootnote#1{\insert\footins\bgroup
\baselineskip10pt\footfont
\interlinepenalty=\interfootnotelinepenalty
\floatingpenalty=20000
\splittopskip=\ht\strutbox \boxmaxdepth=\dp\strutbox
\leftskip=24pt \rightskip=\z@skip
\parindent=12pt \parfillskip=0pt plus 1fil
\spaceskip=\z@skip \xspaceskip=\z@skip
\Textindent{$#1$}\footstrut\futurelet\next\fo@t}
\def\Textindent#1{\noindent\llap{#1\enspace}\ignorespaces}
\def\footnote#1{\attach{#1}\vfootnote{#1}}%

\def\foot{\attach\footsymbolgen\vfootnote{\footsymbol}}
\let\footsymbol=\star
\newcount\lastf@@t           \lastf@@t=-1
\newcount\footsymbolcount    \footsymbolcount=0
\def\footsymbolgen{\relax\footsym
\global\lastf@@t=\pageno\footsymbol}
\def\footsym{\ifnum\footsymbolcount<0
\global\footsymbolcount=0\fi
{\iffrontpage \else \advance\lastf@@t by 1 \fi
\ifnum\lastf@@t<\pageno \global\footsymbolcount=0
\else \global\advance\footsymbolcount by 1 \fi }
\ifcase\footsymbolcount \fd@f\star\or
\fd@f\dagger\or \fd@f\ast\or
\fd@f\ddagger\or \fd@f\natural\or
\fd@f\diamond\or \fd@f\bullet\or
\fd@f\nabla\else \fd@f\dagger
\global\footsymbolcount=0 \fi }
\def\fd@f#1{\xdef\footsymbol{#1}}
\def\space@ver#1{\let\@sf=\empty \ifmmode #1\else \ifhmode
\edef\@sf{\spacefactor=\the\spacefactor}
\unskip${}#1$\relax\fi\fi}
\def\attach#1{\space@ver{\strut^{\mkern 2mu #1} }\@sf\ }
%
\newif\ifnref
\def\rrr#1#2{\relax\ifnref\nref#1{#2}\else\ref#1{#2}\fi}

\nreffalse
\def\refout{\listrefs}
%
\def\eqn#1{\xdef #1{(\secsym\the\meqno)}
\writedef{#1\leftbracket#1}%
\global\advance\meqno by1\eqno#1\eqlabeL#1}
\def\eqnalign#1{\xdef #1{(\secsym\the\meqno)}
\writedef{#1\leftbracket#1}%
\global\advance\meqno by1#1\eqlabeL{#1}}
%
\def\chap#1{\newsec{#1}}
\def\chapter#1{\chap{#1}}
\def\sect#1{\subsec{#1}}
\def\section#1{\sect{#1}}
\def\\{\ifnum\lastpenalty=-10000\relax
\else\hfil\penalty-10000\fi\ignorespaces}
\def\note#1{\leavevmode%
\edef\@@marginsf{\spacefactor=\the\spacefactor\relax}%
\ifdraft\strut\vadjust{%
\hbox to0pt{\hskip\hsize\hskip.05in%
\vbox to0pt{\vskip-\dp\strutbox%
\sevenrm\baselineskip=10pt plus 1pt minus 1pt%
\ifx\answ\bigans\hsize=.9in\else\hsize=.4in\fi%
\tolerance=5000 \hbadness=5000%
\leftskip=0pt \rightskip=0pt \everypar={}%
\raggedright\parskip=0pt \parindent=0pt%
\vskip-\ht\strutbox\noindent\strut#1\par%
\vss}\hss}}\fi\@@marginsf\kern-.01cm}
\def\titlepage{%
\frontpagetrue\nopagenumbers\abstractfont%
\hsize=\hstitle\rightline{\vbox{\baselineskip=10pt%
{\abstractfont\pubnum}}}\pageno=0}
\frontpagefalse
\def\pubnum{}
\def\pdate{\number\month/\number\yearltd}
\def\makefootline{\iffrontpage\vskip .27truein
\line{\the\footline}
\vskip -.1truein\line{\pdate\hfil}
\fi}
\def\title#1{\vskip .7truecm\titlestyle{\titleft #1}}
\def\titlestyle#1{\par\begingroup \interlinepenalty=9999
\leftskip=0.02\hsize plus 0.23\hsize minus 0.02\hsize
\rightskip=\leftskip \parfillskip=0pt
\hyphenpenalty=9000 \exhyphenpenalty=9000
\tolerance=9999 \pretolerance=9000
\spaceskip=0.333em \xspaceskip=0.5em
\noindent #1\par\endgroup }
\def\autskip{\ifx\answ\bigans\vskip.5truecm\else\vskip.1cm\fi}
\def\author#1{\vskip .7in \centerline{#1}}
\def\andauthor#1{\autskip
\centerline{\it and} \autskip\centerline{#1}}
\def\address#1{\ifx\answ\bigans\vskip.2truecm
\else\vskip.1cm\fi{\it \centerline{#1}}}
\def\abstract#1{\vskip .5in\vfil\centerline
{\bf Abstract}\penalty1000
{{\smallskip\ifx\answ\bigans\leftskip 2pc \rightskip 2pc
\else\leftskip 5pc \rightskip 5pc\fi
\noindent\abstractfont \baselineskip=12pt
{#1} \smallskip}}
\penalty-1000}
\def\endpage{\tenpoint\supereject\global\hsize=\hsbody%
\frontpagefalse\footline={\hss\tenrm\folio\hss}}
%
\def\CERN{\address{CERN, Geneva, Switzerland}}
\def\inbar{\vrule height1.5ex width.4pt depth0pt}
\def\IC{\relax\,\hbox{$\inbar\kern-.3em{\mss C}$}}
\def\IF{\relax{\rm I\kern-.18em F}}
\def\IH{\relax{\rm I\kern-.18em H}}
\def\II{\relax{\rm I\kern-.17em I}}
\def\IN{\relax{\rm I\kern-.18em N}}
\def\IP{\relax{\rm I\kern-.18em P}}
\def\IQ{\relax\,\hbox{$\inbar\kern-.3em{\rm Q}$}}
\def\IR{\relax{\rm I\kern-.18em R}}
\def\ZZ{\relax{\hbox{\mss Z\kern-.42em Z}}}
\def\nup#1({Nucl.\ Phys.\ $\us {B#1}$\ (}
\def\plt#1({Phys.\ Lett.\ $\us  {B#1}$\ (}
\def\plb#1({Phys.\ Lett.\ $\us  {#1B}$\ (}
\def\cmp#1({Comm.\ Math.\ Phys.\ $\us  {#1}$\ (}
\def\prp#1({Phys.\ Rep.\ $\us  {#1}$\ (}
\def\prl#1({Phys.\ Rev.\ Lett.\ $\us  {#1}$\ (}
\def\prv#1({Phys.\ Rev. $\us  {#1}$\ (}
\def\und#1({            $\us  {#1}$\ (}
\def\tit#1,{{\it #1},\ }
%

\def\bar{\overline}
\def\us#1{\bf{#1}}

\def\Coe#1.#2.{{#1\over #2}}

\def\coe#1.#2.{\relax{\textstyle {#1 \over #2}}\displaystyle}

\def\notin{\hbox{{$\in$}\kern-.51em\hbox{/}}}

\catcode`\@=12
%
\newbox\hdbox%
\newcount\hdrows%
\newcount\multispancount%
\newcount\ncase%
\newcount\ncols
\newcount\nrows%
\newcount\nspan%
\newcount\ntemp%
\newdimen\hdsize%
\newdimen\newhdsize%
\newdimen\parasize%
\newdimen\spreadwidth%
\newdimen\thicksize%
\newdimen\thinsize%
\newdimen\tablewidth%
\newif\ifcentertables%
\newif\ifendsize%
\newif\iffirstrow%
\newif\iftableinfo%
\newtoks\dbt%
\newtoks\hdtks%
\newtoks\savetks%
\newtoks\tableLETtokens%
\newtoks\tabletokens%
\newtoks\widthspec%
%
%
%
%
\tableinfotrue%
\catcode`\@=11
%
%
\def\tstrut{\vrule height3.1ex depth1.2ex width0pt}%
\def\and{\char`\&}
\def\tablerule{\noalign{\hrule height\thinsize depth0pt}}%
\thicksize=1.5pt
\thinsize=0.6pt
\def\thickrule{\noalign{\hrule height\thicksize depth0pt}}%
\def\ctr#1{\hfil\ #1\hfil}%
%
%
%
%
\tablewidth=-\maxdimen%
\spreadwidth=-\maxdimen%
\def\tabskipglue{0pt plus 1fil minus 1fil}%
%
%
\centertablestrue%
%
%
%
%
\parasize=4in%
\gdef\ARGS{########}
\gdef\headerARGS{####}
\def\@mpersand{&}
{\catcode`\|=13
\gdef\letbarzero{\let|0}
\gdef\letbartab{\def|{&&}}%
\gdef\letvbbar{\let\vb|}%
}
{\catcode`\&=4
\def\ampskip{&\omit\hfil&}
\catcode`\&=13
\let&0
\xdef\letampskip{\def&{\ampskip}}%
\gdef\letnovbamp{\let\novb&\let\tab&}
}
\def\begintable{
   \begingroup%
   \catcode`\|=13\letbartab\letvbbar%
   \catcode`\&=13\letampskip\letnovbamp%
   \def\multispan##1{
      \omit \mscount##1%
      \multiply\mscount\tw@\advance\mscount\m@ne%
      \loop\ifnum\mscount>\@ne \sp@n\repeat%
   }
   \def\|{%
      &\omit\widevline&%
   }%
   \ruledtable
}
\long\def\ruledtable#1\endtable{%
%
%
%
   \offinterlineskip
   \tabskip 0pt
   \def\widevline{\vrule width\thicksize}
   \def\endrow{\@mpersand\omit\hfil\crnorm\@mpersand}%
   \def\crthick{\@mpersand\crnorm\thickrule\@mpersand}%
   \def\crthickneg##1{\@mpersand\crnorm\thickrule
          \noalign{{\skip0=##1\vskip-\skip0}}\@mpersand}%
   \def\crnorule{\@mpersand\crnorm\@mpersand}%
   \def\crnoruleneg##1{\@mpersand\crnorm
          \noalign{{\skip0=##1\vskip-\skip0}}\@mpersand}%
   \let\nr=\crnorule
   \def\endtable{\@mpersand\crnorm\thickrule}%
   \let\crnorm=\cr
%
%
   \edef\cr{\@mpersand\crnorm\tablerule\@mpersand}%
   \def\crneg##1{\@mpersand\crnorm\tablerule
          \noalign{{\skip0=##1\vskip-\skip0}}\@mpersand}%
   \let\ctneg=\crthickneg
   \let\nrneg=\crnoruleneg
   \the\tableLETtokens
%
%
   \tabletokens={&#1}
%
%
   \countROWS\tabletokens\into\nrows%
   \countCOLS\tabletokens\into\ncols%
%
%
   \advance\ncols by -1%
   \divide\ncols by 2%
   \advance\nrows by 1%
%
%
   \iftableinfo %
      \immediate\write16{[Nrows=\the\nrows, Ncols=\the\ncols]}%
   \fi%
%
%
   \ifcentertables
      \ifhmode \par\fi
      \hbox to \hsize{
      \hss
   \else %
      \hbox{%
   \fi
      \vbox{%
         \makePREAMBLE{\the\ncols}
         \edef\next{\preamble}
         \let\preamble=\next
         \makeTABLE{\preamble}{\tabletokens}
      }
      \ifcentertables \hss}\else }\fi
   \endgroup
   \tablewidth=-\maxdimen
   \spreadwidth=-\maxdimen
}
\def\makeTABLE#1#2{
   {
   \let\ifmath0
   \let\header0
   \let\multispan0
%
%
   \ncase=0%
   \ifdim\tablewidth>-\maxdimen \ncase=1\fi%
   \ifdim\spreadwidth>-\maxdimen \ncase=2\fi%
   \relax
%
   \ifcase\ncase %
      \widthspec={}%
   \or %
      \widthspec=\expandafter{\expandafter t\expandafter o%
                 \the\tablewidth}%
   \else %
      \widthspec=\expandafter{\expandafter s\expandafter p\expandafter r%
                 \expandafter e\expandafter a\expandafter d%
                 \the\spreadwidth}%
   \fi %
   \xdef\next{
      \halign\the\widthspec{%
      #1
      \noalign{\hrule height\thicksize depth0pt}
      \the#2\endtable
%
      }
   }
   }
   \next
}
\def\makePREAMBLE#1{
   \ncols=#1
   \begingroup
   \let\ARGS=0
   \edef\xtp{\widevline\ARGS\tabskip\tabskipglue%
   &\ctr{\ARGS}\tstrut}
   \advance\ncols by -1
   \loop
      \ifnum\ncols>0 %
      \advance\ncols by -1%
      \edef\xtp{\xtp&\vrule width\thinsize\ARGS&\ctr{\ARGS}}%
   \repeat
   \xdef\preamble{\xtp&\widevline\ARGS\tabskip0pt%
   \crnorm}
   \endgroup
}
\def\countROWS#1\into#2{
   \let\countREGISTER=#2%
   \countREGISTER=0%
   \expandafter\ROWcount\the#1\endcount%
}%
\def\ROWcount{%
   \afterassignment\subROWcount\let\next= %
}%
\def\subROWcount{%
   \ifx\next\endcount %
      \let\next=\relax%
   \else%
      \ncase=0%
      \ifx\next\cr %
         \global\advance\countREGISTER by 1%
         \ncase=0%
      \fi%
      \ifx\next\endrow %
         \global\advance\countREGISTER by 1%
         \ncase=0%
      \fi%
      \ifx\next\crthick %
         \global\advance\countREGISTER by 1%
         \ncase=0%
      \fi%
      \ifx\next\crnorule %
         \global\advance\countREGISTER by 1%
         \ncase=0%
      \fi%
      \ifx\next\crthickneg %
         \global\advance\countREGISTER by 1%
         \ncase=0%
      \fi%
      \ifx\next\crnoruleneg %
         \global\advance\countREGISTER by 1%
         \ncase=0%
      \fi%
      \ifx\next\crneg %
         \global\advance\countREGISTER by 1%
         \ncase=0%
      \fi%
      \ifx\next\header %
         \ncase=1%
      \fi%
      \relax%
      \ifcase\ncase %
         \let\next\ROWcount%
      \or %
         \let\next\argROWskip%
      \else %
      \fi%
   \fi%
   \next%
}
\def\counthdROWS#1\into#2{%
\dvr{10}%
   \let\countREGISTER=#2%
   \countREGISTER=0%
\dvr{11}%
\dvr{13}%
   \expandafter\hdROWcount\the#1\endcount%
\dvr{12}%
}%
\def\hdROWcount{%
   \afterassignment\subhdROWcount\let\next= %
}%
\def\subhdROWcount{%
   \ifx\next\endcount %
      \let\next=\relax%
   \else%
      \ncase=0%
      \ifx\next\cr %
         \global\advance\countREGISTER by 1%
         \ncase=0%
      \fi%
      \ifx\next\endrow %
         \global\advance\countREGISTER by 1%
         \ncase=0%
      \fi%
      \ifx\next\crthick %
         \global\advance\countREGISTER by 1%
         \ncase=0%
      \fi%
      \ifx\next\crnorule %
         \global\advance\countREGISTER by 1%
         \ncase=0%
      \fi%
      \ifx\next\header %
         \ncase=1%
      \fi%
\relax%
      \ifcase\ncase %
         \let\next\hdROWcount%
      \or%
         \let\next\arghdROWskip%
      \else %
      \fi%
   \fi%
   \next%
}%
{\catcode`\|=13\letbartab
\gdef\countCOLS#1\into#2{%
   \let\countREGISTER=#2%
   \global\countREGISTER=0%
   \global\multispancount=0%
   \global\firstrowtrue
   \expandafter\COLcount\the#1\endcount%
   \global\advance\countREGISTER by 3%
   \global\advance\countREGISTER by -\multispancount
}%
\gdef\COLcount{%
   \afterassignment\subCOLcount\let\next= %
}%
{\catcode`\&=13%
\gdef\subCOLcount{%
   \ifx\next\endcount %
      \let\next=\relax%
   \else%
      \ncase=0%
      \iffirstrow
         \ifx\next& %
            \global\advance\countREGISTER by 2%
            \ncase=0%
         \fi%
         \ifx\next\span %
            \global\advance\countREGISTER by 1%
            \ncase=0%
         \fi%
         \ifx\next| %
            \global\advance\countREGISTER by 2%
            \ncase=0%
         \fi
         \ifx\next\|
            \global\advance\countREGISTER by 2%
            \ncase=0%
         \fi
         \ifx\next\multispan
            \ncase=1%
            \global\advance\multispancount by 1%
         \fi
         \ifx\next\header
            \ncase=2%
         \fi
         \ifx\next\cr       \global\firstrowfalse \fi
         \ifx\next\endrow   \global\firstrowfalse \fi
         \ifx\next\crthick  \global\firstrowfalse \fi
         \ifx\next\crnorule \global\firstrowfalse \fi
         \ifx\next\crnoruleneg \global\firstrowfalse \fi
         \ifx\next\crthickneg  \global\firstrowfalse \fi
         \ifx\next\crneg       \global\firstrowfalse \fi
      \fi
\relax
      \ifcase\ncase %
         \let\next\COLcount%
      \or %
         \let\next\spancount%
      \or %
         \let\next\argCOLskip%
      \else %
      \fi %
   \fi%
   \next%
}%
\gdef\argROWskip#1{%
   \let\next\ROWcount \next%
}
\gdef\arghdROWskip#1{%
   \let\next\ROWcount \next%
}
\gdef\argCOLskip#1{%
   \let\next\COLcount \next%
}
}
}
\def\spancount#1{
   \nspan=#1\multiply\nspan by 2\advance\nspan by -1%
   \global\advance \countREGISTER by \nspan
   \let\next\COLcount \next}%
\def\dvr#1{\relax}%
\def\header#1{%
\dvr{1}{\let\cr=\@mpersand%
\hdtks={#1}%
\counthdROWS\hdtks\into\hdrows%
\advance\hdrows by 1%
\ifnum\hdrows=0 \hdrows=1 \fi%
\dvr{5}\makehdPREAMBLE{\the\hdrows}%
\dvr{6}\getHDdimen{#1}%
{\parindent=0pt\hsize=\hdsize{\let\ifmath0%
\xdef\next{\valign{\headerpreamble #1\crnorm}}}\dvr{7}\next\dvr{8}%
}%
}\dvr{2}}
\def\makehdPREAMBLE#1{
\dvr{3}%
\hdrows=#1
{
\let\headerARGS=0%
\let\cr=\crnorm%
\edef\xtp{\vfil\hfil\hbox{\headerARGS}\hfil\vfil}%
\advance\hdrows by -1
\loop
\ifnum\hdrows>0%
\advance\hdrows by -1%
\edef\xtp{\xtp&\vfil\hfil\hbox{\headerARGS}\hfil\vfil}%
\repeat%
\xdef\headerpreamble{\xtp\crcr}%
}
\dvr{4}}
\def\getHDdimen#1{%
\hdsize=0pt%
\getsize#1\cr\end\cr%
}
\def\getsize#1\cr{%
\endsizefalse\savetks={#1}%
\expandafter\lookend\the\savetks\cr%
\relax \ifendsize \let\next\relax \else%
\setbox\hdbox=\hbox{#1}\newhdsize=1.0\wd\hdbox%
\ifdim\newhdsize>\hdsize \hdsize=\newhdsize \fi%
\let\next\getsize \fi%
\next%
}%
\def\lookend{\afterassignment\sublookend\let\looknext= }%
\def\sublookend{\relax%
\ifx\looknext\cr %
\let\looknext\relax \else %
   \relax
   \ifx\looknext\end \global\endsizetrue \fi%
   \let\looknext=\lookend%
    \fi \looknext%
}%
%
%
\def\tablelet#1{%
   \tableLETtokens=\expandafter{\the\tableLETtokens #1}%
}%
\catcode`\@=12
%

%
\def\OFFSET{\hoffset=6.pt\voffset=40.pt}

\OFFSET

\def\WIT{\rrr\WIT{E. Witten, \prv44 (1991) 314.}}

\def\WZW{\rrr\WZW{E. Witten, Comm. Math. Phys. {\bf 92} (1984) 455;
E. Witten, \nup223 (1983) 422;
K. Bardacki, E. Rabinovici and B. Saering, \nup301 (1988) 151;
D. Karabali and H.J. Schnitzer, \nup 329 (1990) 649.}}

\def\DLP{\rrr\DLP{L. Dixon, J. Lykken and M. Peskin, \nup325 (1989)
325;
I. Bars, \nup334 (1990) 125; I. Bars and D. Nemeschansky, \nup348
(1991) 89.}}

\def\BUSCH{\rrr\BUSCH{T. Buscher, \plt194 (1987) 59, \plt201 (1988)
466;
L.E. Ib\'a\~nez, D. L\"ust, F. Quevedo and
S. Theisen, unpublished notes (1990);
E. Smith and J. Polchinski, \plt 263 (1991) 59;
G. Veneziano, \plt265 (1991) 287;
A.A. Tseytlin, Mod. Phys. Lett. {\bf A6} (1991) 1721.}}

\def\KIR{\rrr\KIR{E.B. Kiritsis, Mod. Phys. Lett. {\bf A6} (1991) 2871.}}

\def\GIV{\rrr\GIV{A. Giveon, Mod. Phys. Lett. {\bf A6} (1991) 2843.}}

\def\RAB{\rrr\RAB{S. Elitzur, A. Forge and E. Rabinovici, \nup359
(1991) 581; G. Mandal, A.M. Sengupta and S.R. Wadia,
Mod. Phys. Lett. {\bf A6} (1991) 1685.}}

\def\DIVER{\rrr\DIVER{R. Dijgkraaf, E. Verlinde and H. Verlinde,
\nup371 (1992) 269.}}

\def\CFMP{\rrr\CFMP{C. Callan, D. Friedan, E. Martinec and M. Perry,
\nup262 (1985) 593.}}

\def\MUL{\rrr\MUL{M. Muller, \nup337 (1990) 37.}}

\def\HOR{\rrr\HOR{P. Horava, \plt278 (1992) 101.}}

\def\ANT{\rrr\ANT{I. Antoniadis, C. Bachas, J. Ellis and D.V. Nanopoulos,
\plt211 (1988) 393, \nup328 (1989) 117.}}

\def\MYE{\rrr\MYE{R.C. Myers, \plt199 (1987) 371; A.A. Tseytlin and
C. Vafa, \nup372 (1992) 443; A.A. Tseytlin, preprint
DAMTP-91-37; A.A. Tseytlin, preprint DAMPT-92-15;
K. Behrndt, preprint DESY 92-055.}}

\def\GIQU{\rrr\GIQU{
J.H. Horne and G. Horowitz, \nup368 (1992) 444;
I. Bars and Sfetsos, \plt277 (1992) 269;
D. Gershon, preprint TAUP-1937-91 (1991);
J. Ellis, N.E. Mavromatos and D.V. Nanopoulos, \plt278 (1992) 246;
P. Ginsparg and F. Quevedo, preprint LA-UR-92-640
(1992).}}

\def\HWANG{\rrr\HWANG{S. Hwang, \nup354 (1991) 100; H. Hennigson,
S. Hwang and P. Roberts, \plt267 (1991) 350.}}

\def\HAEL{\rrr\HAEL{
S.W. Hawking and G.F.R. Ellis, {\it ``The Large Scale Structure of
Space-Time'',} Cambridge University Press, Cambridge, 1973.}}

\def \ANFEKOU{\rrr\ANFEKOU{I. Antoniadis, S. Ferrara and C. Kounnas,
preprint LPTENS 91-30; I. Antoniadis and C. Kounnas, preprint
LPTENS 91-31.}}

%


\def\pubnum{
\hbox{CERN-TH.6494/92}\hbox{LPTENS 92-16}}

\def\pdate{May 1992}

\titlepage
\title
{Cosmological String Backgrounds from Gauged WZW Models}
\vskip-.8cm
\author{{\bf Costas Kounnas}}
\vskip.2cm
\centerline{\it Ecole Normale Sup\'erieure, Paris, France}
\andauthor{{\bf Dieter L\"ust}\foot{Heisenberg Fellow}}
\CERN
\vskip-2.8 cm
\abstract{We discuss the four-dimensional target-space interpretation of
bosonic strings based on gauged WZW models, in particular of those
based on the non-compact coset space $SL(2,{\bf R})\times SO(1,1)^2
/SO(1,1)$.
We show that these theories lead,
apart from the recently
broadly discussed black-hole type of backgrounds,
to cosmological string
backgrounds, such as an expanding Universe.
Which of the two cases is realized depends on the sign of the level
of the corresponding Kac-Moody algebra.
We discuss various aspects of these new cosmological
string backgrounds.}

\vskip2.5cm
\endpage
%
\nopagenumbers
\null\vfill\eject
\advance\pageno -1

\def\makefootline{
\vskip.5cm\line{\hss \tenrm $-$ \folio\ $-$ \hss}}

The investigation of strings in curved space-time backgrounds might
provide us with some important insight into questions about
quantum gravity. Recently, a very intriguing relation between
black-hole type of backgrounds \RAB\ in two dimensions and
gauged Wess-Zumino-Witten (WZW) models \WZW\ based on the
non-compact coset space $SL(2,{\bf R})/SO(1,1)$
was discovered \WIT. Subsequently, this discussion was put
forward to higher dimensions \GIQU. All these examples show
a singularity in forward times (black-holes, black branes, etc.).
In this note we shall argue
that the same non-compact cosets, which give rise to
black-holes,
also lead to cosmological string backgrounds, namely
to an expanding Universe. (Cosmological string backgrounds were
discussed before in \MYE,\ANT,\MUL.)
In this case, the black-hole singularity becomes a singular
surface hidden behind the light-cone, which can never be met in
future times when one travels inside the light-cone.
Even our cosmological string solution is derived from an exact
conformal field theory in the semi-classical approximation,
we think that it also serves to be
an interesting solution in its own right
to Einstein's
equations coupled to a non-constant dilaton matter field.

The gauged WZW model based on the coset $G/H$ is described by the
action
$$\eqalign{S=&{k\over 4\pi}\int {\rm d}^2z{\rm tr}(g^{-1}\partial \
g^{-1}\bar\partial g)-{k\over 12\pi}\int_B{\rm tr}(g^{-1}dg\wedge g^{-1}
dg\wedge g^{-1}dg)\cr
&+{k\over2\pi}\int{\rm d}^2{\rm tr}(A\bar\partial gg^{-1}-\bar Ag^{-1}
\partial g-g^{-1}Ag\bar A),\cr}\eqn\wzwaction
$$
where the boundary of $B$ is the 2D worldsheet,
$g$ is a group element of the group $G$, and $A$ are the gauge fields
of $H$ transforming as $A\rightarrow h_L^{-1}(A+\partial )h_L$; $k$ is
the level of the Kac--Moody algebra for the group $G$. To be specific,
we want to discuss a four-dimensional target-space
based on the non-compact coset $SL(2,{\bf R})\times SO(1,1)^2/SO(1,1)$.
The central charge of this coset CFT (non-compact coset CFT's were
discussed in \DLP)
is given by
$$c=4+{6\over k-2},\eqn\central
$$
where the $k$, being a real number, is the level of the non-compact
$SL(2,{\bf R})$ Kac--Moody algebra.
For the case that the string theory
is entirely given by this coset CFT, the condition $c=26$ implies
$k=25/11$. However, we
want to study the semiclassical limit $k\rightarrow\pm
\infty$. Therefore we prefer to couple the $SL(2,{\bf R})\times SO(1,1)^2/
SO(1,1)$ coset CFT to an internal CFT with central charge
$$c_{\rm int}=
22-\delta c,\qquad
\delta c={6\over k-2}.\eqn\centralch
$$
Now we parametrize the
group element of $SL(2,{\bf R})$ as
$g=\pmatrix{u&a\cr -b&v\cr}$ with $uv+ab=1$.
Finally we assume for simplicity that $H=SO(1,1)$
is entirely inside $SL(2,{\bf R})$. Performing a vector-like
gauging, $a=\pm b$, the action \wzwaction\ can be identified with
a $\sigma$-model action of the form
$$S=\int{\rm d}^2zG_{MN}^\sigma
\partial X^M\bar\partial X^N.\eqn\sigmaaction
$$
In the semiclassical approximation $k\rightarrow\pm\infty$,
the corresponding four-dimensional $\sigma$-model metric is given as
$${\rm d}s^2_\sigma=-k{{\rm d}u{\rm d}v \over 1-uv}+{\rm d}x_2^2
+{\rm d}x_3^2.\eqn\sigmametric
$$
We recognize that the signature of the two-dimensional part of
this metric depends just on the sign of the level $k$ of the
$SL(2,{\bf R})$ Kac--Moody algebra. In fact, for $k\rightarrow +\infty$
($\delta c>0$)
space-time possesses a singularity in futures times $\tau=u+v$.
In this
case, $u$ and $v$  are Kruskal-like coordinates of a black-hole metric;
more precisely the metric \sigmametric\ has the causal structure
of a four-dimensional black-brane (see figure 1).
The singularity at $uv=1$ originates from the fixed points
of the modded vector gauge symmetry $H$ at this curve.
However for $k\rightarrow -\infty$ ($\delta c<0$)
the causal structure is completely different. Specifically the
causal structure for negative $k$ is obtained from the black-hole case
by a $90^0$ rotation of the two-dimensional $u,v$-plane (see figure 2).
Now one has  a forward light-cone with the
singularity behind it.
(Of course, the
two  cases $k\rightarrow\pm\infty$ are not analytic continuations
of each other.) As we will discuss in the following, this new class
of string backgrounds with negative $k$
describes an expanding Universe with singularity
outside the visible horizon.

The four-dimensional effective string action  for
the gravitational  and dilaton $\Phi$
background fields has the following form:
$$S_{\rm eff}=\int{\rm d}^4x\sqrt{G^\sigma}e^\Phi(R^\sigma -(D\Phi)^2
+\Lambda),\qquad \Lambda={2\delta c\over 3}.\eqn\effeaction
$$
Here $R^\sigma$ is the curvature scalar derived from the metric
$G_{MN}^\sigma$ in the
$\sigma$-model frame. It is not
difficult to show that the $\sigma$-model metric eq.\sigmametric\
together with the dilaton field
$$\Phi=\log(1-uv)\eqn\dilaton
$$
satisfy the equations of motion \CFMP
$$\eqalign{&R_{MN}^\sigma+D_MD_N\Phi=0,\cr
&R^\sigma+(D_M\Phi)^2+2D_MD^M\Phi=\Lambda,\cr}\eqn\sigmaeq
$$
for both signs of $k$.
(In fact, one has to replace $k$ by $k-2=6/\delta c$ in eq.\sigmametric,
which is allowed in the large $k$ limit.)

To get contact with standard gravity theory, which possesses
a canonical Einstein term, one has to perform a Weyl rescaling
of the $\sigma$-model metric eq.\sigmametric\ by the exponential
of the dilaton field.\foot{There is no Weyl rescaling of the metric
in two space-time dimensions.}
This Weyl rescaling however does not change the
causal structure of the theory.
Specifically, the metric in the Einstein
frame is given as
$${\rm d}s^2=e^\Phi{\rm d}s^2_\sigma={\rm d}u{\rm d}v
+(1-uv)({\rm d}x_2^2+{\rm d}x_3^2).\eqn\einsteinmetric
$$
Here we have focused on the
cosmological case with $k\rightarrow -\infty$,
where we have absorbed the level $k$ in the metric. The effective
action now has the form
$$S_{\rm eff}=\int{\rm d}^4x\sqrt G\biggl({1\over 2}R-{1\over 4}D_M\Phi
D^M\Phi-V(\Phi)\biggr) ,\qquad V(\Phi)=2e^{-\Phi}.\eqn\einsteinaction
$$

The metric eq.\einsteinmetric\ in the Kruskal-type of coordinates
$u$ and $v$ leads to a space-time singularity at $uv=1$.
This can been seen by computing the corresponding Ricci tensor:
$$R_{uv}={uv-2\over 2(1-uv)^2}.\eqn\ricci
$$
Second, there is the horizon at $uv=0$. Thus we see
that,
introducing the proper time $\tau=u-v$, there is  no singularity
for future times $\tau$ inside the light-cone $uv<0$
(region I in figure
2). The singularity is hidden behind the horizon. However signals in
regions II, III
may hit the singularity, and the singularity may also send signals
through regions II, III into region I.

Now we want to show that inside the singularity free
region I we have in fact an expanding Universe with metric similar to the
Robertson-Walker metric. For this purpose it is useful to
introduce coordinates which cover exactly the region I,
namely\foot{Alternatively, one can change the coordinates in the
following way: $u=e^{x_1}\sqrt{t-1}$, $v=-e^{-x_1}\sqrt{t-1}$ in region I
and $u=\pm e^{x_1}\sqrt{1-t}$, $v=\pm e^{-x_1}\sqrt{1-t}$ in regions II,
III.
Then the metric takes the form ${\rm d}s^2=-{1\over 4(t-1)}{\rm d}t^2
+(t-1){\rm d}x_1^2+t({\rm d}x_2^2+{\rm d}x_3^2)$. In region I this
metric again  leads to an expanding Universe, isotropic for large $t$.}
$$u=e^{x_1}t,\qquad v=-e^{-x_1}t.\eqn\change
$$
(The coordinates $t$ and $x_1$ are of course not geodesically complete;
the singularity occurs for imaginary times $t$, $t^2<-1$.)
It follows that curves of constant times $t$ correspond in figure 2 to
hyperbolae $uv=-t^2$, whereas the curves of constant $x_1$ are given
by the straight lines $u/v=-e^{2x_1}$ (see figure 2).
In these coordinates
the metric now looks like
$${\rm d}s^2=-{\rm d}t^2+t^2{\rm d}x_1^2
+(1+t^2)({\rm d}x_2^2+{\rm d}x_3^2),\eqn\newmetric
$$
and the dilaton has the form
$$\Phi(t)=\log(1+t^2).\eqn\newdil
$$
Clearly, the metric \newmetric\ describes an expanding Universe in region
I with two different scale factors $R_1(t)=t$, $R_{2,3}(t)=\sqrt{1+t^2}$,
where $t$ is the cosmological time coordinate.
For small $t$, the Universe expands unisotropically. However, for
large times, one approaches an isotropic, linear expansion of the
Friedmann-Robertson-Walker type with $R_i(t)=t$ ($i=1,2,3$).
Thus we see that for large $t$ this background, derived from the coset
CFT $SL(2,{\bf R})\times SO(1,1)^2/SO(1,1)$, asymptotically
approaches the linearly expanding Universe considered in ref.\ANT,
which is based on a free scalar CFT plus a dilaton of the form
$\Phi=2\log t$.
However our solution has no initial singularity at $t=0$ in contrast to
the standard isotropic Robertson-Walker Universe.
(The Ricci tensor in the coordinates \change\ takes
the form $R_{tt}={2\over (1+t^2)^2}$, $R^i_j=-{2\over 1+t^2}\delta^i_j$.)

Let us briefly derive the energy momentum tensor of the dilaton matter
field.
Specifically consider the
classical Einstein equations
$$R_{MN}-{1\over 2}G_{MN}R=-T_{MN}.\eqn\einsteinequ
$$
The corresponding
energy-momentum tensor from the dilaton matter field has
the form
$$T_{MN}={1\over 2}D_M\Phi D_N\Phi-G_{MN}\biggl({1\over 4}G^{PQ}D_P\Phi
D_Q\Phi+V(\Phi)\biggr) .\eqn\energymom
$$
Then we obtain, with $\Phi=\log(1+t^2)$ and $V(\Phi)={2\over 1+t^2}$,
that
$$\eqalign{
T_{tt}&={(\partial_t\Phi)^2\over 4}+V=\rho={3t^2+2\over (1+t^2)^2},\cr
T_{ij}&=\biggl({(\partial_t\Phi)^2\over 4}-V\biggr)G_{ij}
=pG_{ij}=-{t^2+2\over (1+t^2)^2}G_{ij}.\cr}\eqn\densitypres
$$
Here $\rho$ is the energy density of the dilaton matter system
and $p$ is its pressure. Now it is easy to see that the quantity
$$\rho+3p=-{4\over (1+t^2)^2}\eqn\rhopi
$$
is negative for all $t$.
It is interesting to observe that the form of $\rho+3p$, being always
negative, violates
an assumption by Hawking and Ellis \HAEL\
on the form
of the matter energy-momentum tensor, which, being satisfied,
would always lead to a singular space-time.
Thus, the absence of an initial
singularity in the cosmological region I can
be understood from this point of view.

It is also interesting to discuss the target space
duality properties of this
cosmological string background. In the gauged WZW model, the
duality operation corresponds to the exchange of
gauging the axial $SO(1,1)$ subgroup of $SL(2,{\bf R})$
instead of gauging the vector-like subgroup $SO(1,1)$ \KIR.
Seen from the target space point of view it is the exchange of the regions
I and IV in figures 1 and 2, whereas the regions II and III
are mapped onto
themselves \GIV,\DIVER. In fact, one can show explicitly \BUSCH\
how the metric \sigmametric\ is transformed under the
duality transformation to a new metric, which also satisfies the field
equations \sigmaeq\
after a suitable redefinition of the dilaton field. For this purpose
it is convenient
to introduce another set of coordinates like
$$u=e^{x_1}{\rm sinh}x_0,\qquad v=-e^{-x_1}{\rm sinh}x_0.\eqn\coordtwo
$$
Then the cosmological string background of region I
in the sigma-model frame has
the form\foot{This metric is exactly of the form of the cosmological
backgrounds discussed in \MUL.}
$${\rm d}s^2_\sigma=-{\rm d}x_0^2+{\rm tanh}^2x_0{\rm d}x_1^2
+{\rm d}x_2^2+{\rm d}x_3^2.\eqn\metrictwo
$$
The duality transformation now acts as
$$R_1(x_0)={\rm tanh}x_0\rightarrow {1\over R_1(x_0)}={\rm coth}x_0.
\eqn\duality
$$
Therefore the dual metric just describes region IV with a singularity
at $x_0=0$. The field equations \sigmaeq\ are invariant under
this duality transformation, provided that the dilaton transforms
as
$$\Phi(x_0)=2\log{\rm cosh}x_0\rightarrow \Phi(x_0)+2\log R_1(X_0)
=2\log{\rm sinh}x_0.\eqn\diltrans
$$
Finally, in the coordinates eq.\change, the duality transformation
is expressed as
$$t^2\rightarrow -1-t^2.\eqn\dualt
$$
Thus we see again that the cosmological region I is mapped to
the cosmological region IV, which requires an analytic continuation
to imaginary $t$ values.

Let us discuss briefly some alternative models. For example consider
the gauged WZW model based on the
coset ${SL(2,{\bf R})\over SO(1,1)}
\times {SU(2)\over U(1)}$.\foot{This model is relevant for
curved backgrounds with target-space supersymmetries as discussed
in ref.\ANFEKOU.}
The central
charge of this CFT is given by
$$c={3k_1\over k_1-2}+{3k_2\over k_2+2}-2.\eqn\centralcal
$$
The corresponding
background in the Einstein frame has the following form:
$${\rm d}s^2=-k_1(1+z\bar z )
{\rm d}u{\rm d}v+k_2(1-uv){\rm d}z{\rm d}\bar z.\eqn\sutwo
$$
Here $z$ is a complex parameter of an $SU(2)$ group element,
$z=x_2+ix_3$.
For the black-hole case with positive $k_1$ (the $SU(2)$ Kac--Moody
level $k_2$ should always be a positive integer) one can set
$k_1=k_2+4$. Then one obtains that $c=4$ regardless of the value of
$k_1$.
Thus $\delta c=0$ for all $k_1$
and one could expect that the four-dimensional background
of the semiclassical approximation is valid for all $k_1$.
In addition, since $c_{\rm int}=22$ for all $k_1$, the internal
CFT could be simply given by (compactified) free bosons.

For the cosmological case with negative $k_1$ we cannot set $k_1=k_2+4$.
As an alternative one could couple $SL(2,{\bf R})/SO(1,1)$
with negative $k_1$ to an Euclidean two-dimensional
`black-hole' based on the coset $SL(2,{\bf R})/U(1)$ with positive
Kac--Moody level $k_2$. (A twisted version of
the product of an Euclidean and Minkowskian
black-hole coset CFT was already considered in ref.\HOR.)
The central charge of this model is
$$c={3k_1\over k_1-2}+{3k_2\over k_2-2}-2.\eqn\centralcala
$$
Again one
obtains $\delta c=0$ for $k_1=-k_2+4$. It would be interesting
to study further alternatives to this model.

In summary we have shown that gauged WZW-models  have a completely
different target space interpretation
when one changes the sign of the level of the underlying
non-compact Kac--Moody algebra. For both signs, black-holes as well as
cosmological backgrounds, the metric of the target space leads to
space-time singularities. In the cosmological case the singularity is
hidden behind
the light-cone. However the singularity
could send signals into the light-cone and therefore influence
the expansion of the Universe. This is very similar to the initial
singularity (Big Bang) in the standard Friedmann-Robertson--Walker
Universe.
It is important to stress that, seen from the CFT
point of view, the singularities in the black-hole
as well as in the cosmological
frameworks have exactly the same origin, namely in the existence
of fixed points of the modded (vector) `gauge' symmetry $H$. Therefore
one could expect that the same type of quantum gravity effects
are relevant near the black-hole as well as in the early Universe
at times shortly
after the initial singularity.
These quantum gravity effects should in principle be determined from the
underlying coset CFT. (For considerations in this directions
for the black-hole case, see ref.\DIVER.)
Finally we want to mention that the gauged WZW model based on the
non-compact coset $SL(2,{\bf R})/SO(1,1)$ possesses, for both signs
of $k$\foot{For negative $k$, this coset CFT shows many parallels
to the CFT based on the compact space $SU(2)/U(1)$.}, still
many  problems which are presently not completely
understood. In particular the elimination of the negative-norm
states is unclear.
This question should be addressed together
with the spectrum of the additional part of the CFT. Moreover,
the construction of a modular invariant partition function
is still problematic (for a discussion on this issue see \HWANG).

\bigskip
\bigskip

We like to thank P. Candelas, E. Kiritsis, W. Lerche, N. Mavromatos,
M. Porrati and F. Quevedo
for useful discussions.

\refout
\endpage

\centerline{\bf Figure Captions}
\vskip1cm

\noindent{\bf Figure 1:}
The causal structure of the two-dimensional slice of the black-hole
metric equation \sigmametric\ with positive Kac--Moody level $k$.

\noindent{\bf Figure 2:}
The causal structure of the two-dimensional slice of the cosmological
metric equation \sigmametric\ with negative Kac--Moody level $k$.

\vfill\eject\bye